\documentclass{article}

\usepackage{microtype}
\usepackage{graphicx}
\usepackage{booktabs} 

\usepackage[utf8]{inputenc}
\usepackage{amsfonts}
\usepackage{amsmath}
\usepackage{amsthm}
\usepackage{amssymb}
\usepackage{subcaption}
\usepackage{placeins}
\usepackage{todonotes}
\usepackage{caption}
\usepackage{subcaption}
\usepackage{dblfloatfix}
\usepackage[bottom]{footmisc}

\usepackage[linkcolor=blue]{hyperref}

\usepackage[T1]{fontenc}
\usepackage{listings}
\usepackage{pgfplots}

\pgfplotsset{compat=1.17}
\usepackage[accepted]{icml2020}
\usepackage{soul}  

\begin{document}
\icmltitlerunning{MP3net: coherent, minute-long music generation from raw audio with a simple convolutional GAN}

\twocolumn[
\icmltitle{MP3net: coherent, minute-long music generation from raw audio \\with a simple convolutional GAN}



\icmlsetsymbol{equal}{*}

\begin{icmlauthorlist}
\icmlauthor{Korneel van den Broek}{earth}
\end{icmlauthorlist}

\icmlaffiliation{earth}{}

\icmlcorrespondingauthor{}{korneelvdbroek@gmail.com}

\icmlkeywords{Machine Learning, Generative Models, Music}

\vskip 0.3in
]



\printAffiliationsAndNotice{}  

\begin{abstract}
We present a deep convolutional GAN which leverages techniques from MP3/Vorbis audio compression to produce long, high-quality audio samples with long-range coherence. The model uses a Modified Discrete Cosine Transform (MDCT) data representation, which includes all phase information. Phase generation is hence integral part of the model. We leverage the auditory masking and psychoacoustic perception limit of the human ear to widen the true distribution and stabilize the training process. The model architecture is a deep 2D convolutional network, where each subsequent generator model block increases the resolution along the time axis and adds a higher octave along the frequency axis. The deeper layers are connected with all parts of the output and have the context of the full track. This enables generation of samples which exhibit long-range coherence. We use MP3net to create 95s stereo tracks with a 22kHz sample rate after training for 250h on a single Cloud TPUv2. An additional benefit of the CNN-based model architecture is that generation of new songs is almost instantaneous.
\end{abstract}

\section{Introduction}

\begin{figure*}[t]
    \begin{subfigure}{\textwidth}
      \centering
      \includegraphics[width=\textwidth]{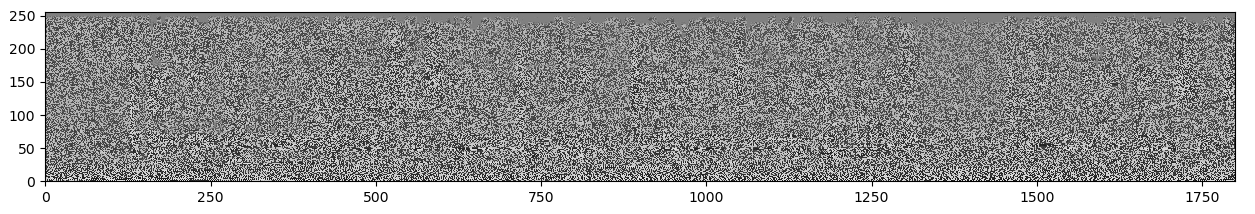}
    \end{subfigure}
    \begin{subfigure}{\textwidth}
      \centering
      \includegraphics[width=\textwidth]{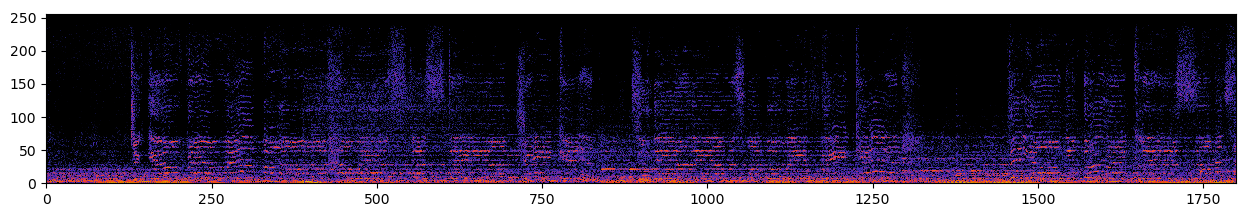}
    \end{subfigure} 
    \caption{On top, the MDCT amplitude representation $A_k(m)$ of a 14 second music sample with a 32,768Hz sample rate. This representation has 256 filter bands on the y-axis ($k: 0 \rightarrow 255$) and shows 1800 blocks on the x-axis. Negative MDCT amplitudes are represented as dark grey and positive amplitudes as light grey. To make the broad range of amplitudes present in the audio signal visible to the eye, the amplitudes are converted to the dB-scale before plotting, otherwise large part of the image would be plain grey as the amplitudes are small.
On the bottom, we show the MDCT spectrogram $A^2_k(m)$ of that same music sample. The intensity is again show on a dB-scale and the color is added as a visual aid to better distinguish the intensity of each pixel in the tensor. Visually, the spectrogram is much easier to interpret compared to the MDCT amplitudes. One can for example clearly make out several harmonics as the set of parallel horizontal lines in the spectrogram.}
    \label{fig:spectrogram}
\end{figure*}

The impressive progress in the field of image generation has resulted in generative models which are able to generate high-resolution images indistinguishable from real pictures \cite{progan, stylegan, selfattention}. The field of generative models for music has long focused on generation of music scores. 
For such models, the final conversion to sound relies on synthesized musical sounds (e.g. MIDI files) to convert the symbolic music into an audio signal. This approach reduces the dimensionality of the audio generation problem, but limits the output to a combination of the set of synthesized sounds.

With the increase in computing power in recent years, generative audio models have moved to generate raw audio directly \cite{wavenet, wavegan, googlespecgan, jukebox}. An high-quality stereo sample of 20 seconds has an uncompressed size of 3.4MiB when sampled at 44.1kHz with a 16-bit resolution. This is similar to the high-resolution 1024x1024 images with 3 8-bit color channels from \citet{progan} which have a 3.0MiB size in its uncompressed representation. \citet{jukebox} show that it is now possible to generate diverse and coherent audio samples several minutes long. However, these generative model require large amounts of computing power.

On the other hand, audio is an integral part of consumer electronic devices, which have much less computing power. This is in large part due to audio compression techniques such as MP3 and Vorbis compressing the audio data into much smaller data formats \cite{mp3explained}. MP3 relies on two key features to achieve a reduction in audio file size, often by as much as 75-93\% \cite{mp3compressionrate}. First, it leverages the psychoacoustic properties of human hearing \cite{pafilteroldarticle}. In particular, if a audio signal is below the threshold for hearing, the human ear will not perceive it. Additionally, certain louder sounds will mask more quiet sounds, even if the more quiet sound would have been audible when played by itself. The second key feature behind MP3 compression is Huffman coding which further decreases the data format size. Huffman coding would not be an easy data format for (convolutional) neural networks to parse, given the codes have variable length. The psychoacoustic properties, on the other hand, allow us to remove a significant amount of data from the audio signal without an audible impact, hence reducing the computing power required to generate raw audio samples with a generative model.

The psychoacoustic properties of the ear are expressed in the frequency domain, so MP3 encoding first converts an audio signal using the modified discrete cosine transformation (MDCT). Compared to other Fourier-related transformations, such as the Short-Time Fourier Transforms (STFT), the MDCT transformation has the added benefit that it is a real transformation. All phase information of the raw audio signal is hence encoded, without the need to use complex numbers, such as in \citet{googlespecgan}, or recreate the phase data with a spectrogram inversion algorithm as used in \citet{melnet}. Additionally, MDCT compacts the original audio signal in fewer amplitude components compared to other Fourier-related transforms.   

In this work, we use the MDCT amplitude as the data representation for raw audio in a deep 2D convolutional Generative Adverserial Network (GAN). In the first layer of the discriminator, we add inaudible psychoacoustic noise to the representation, similar to the quantization noise in MP3 encoding. This widens the limited support of the real distribution, hence stabilizing the training process \cite{stablegan_withnoise2}. 
The architecture of our network is inspired by the ProGAN model \cite{progan} with the noticeable difference that we don't increase/decrease the pixel density along the frequency axis with each successive model-block in the generator/discriminator. Instead, each model-block adds/removes an octave along the frequency axis. 

MP3net does not require spectrogram inversion since the phase information is fully contained in the data representation. By training the top layers of the model with deeper weights frozen, we can eliminate noise and scratchiness in the audio. 
As the deeper layers of the convolutional network have the full context of the entire sample, MP3net lends itself more naturally to long-range coherence. 
Moreover, the model CNN-based architecture allows for almost instantaneous generation of new samples. 

\section{Model}

\subsection{Modified Discrete Cosine Transform (MDCT) amplitudes as data representation}
Given a signal $x(t)$ sampled over $t \in [0, T[$ with sample rate $f_s = 1 / t_0$, we can group the sampled datapoints in $M$ blocks of length $N$: 
\begin{equation} \label{eq:sampling}
  x\Big((mN + n)t_0 \Big) \; \text{with} \; \left\{
  \begin{array}{rcl}
    n & = & 0, \dots, N-1 \\
    m & = & 0, \dots, M-1
  \end{array}
  \right.
\end{equation}
where $T = N M t_0$.

By applying on each pair of consecutive blocks the following linear transformation, we can compute the MDCT amplitudes $A_k(m)$:
\begin{multline}
  A_k(m) = \sum_{n=0}^{2N-1} x\Big((mN+n)t_0\Big) \; w_n  \\
 \cos \left[ \frac{\pi}{N} \left(n + \frac{1}{2} + \frac{N}{2}\right) \left(k + \frac{1}{2} \right) \right] 
\end{multline}
where $k = 0, \dots, N - 1$ is the frequency index corresponding with the following frequency filter band:
\begin{equation} \label{eq:freq}
  [f^0_k, f^0_k + \frac{f_s}{2N} [ = \left[ \frac{f_s}{2N} k, \frac{f_s}{2N} \left( k + 1 \right) \right[
\end{equation}
The discretized window function $w_n$ satisfies
\begin{eqnarray}
  w_n = w_{2N-1-1} \\
  w_n^2 + w_{n+N}^2 = 1
\end{eqnarray}
These latter conditions ensure that the MDCT transformation is invertible. As such, MDCT is a lossless linear transformation of the raw audio signal. Multiple choices exist for the window function, in our implementation we opted for the window function used by the Vorbis project behind the Ogg data format:
\begin{equation}
  w_n = \sin\left(\frac{\pi}{2} \sin^2 \left[\frac{\pi}{2N}\left(n+\frac{1}{2}\right)\right]\right)
\end{equation}

The MDCT is similar to the Short-Time Fourier Transform (STFT) which has been applied in several recent music generation models \cite{wavegan, melnet, googlespecgan},
\begin{equation}
  \tilde{A}_k(m) = \sum_{n=-\infty}^{\infty} x(nT) w_{n-m} \big( \cos kn - i \sin kn \big)
\end{equation}
where $w_n$ is a discretization of a window function with compact support.

Compared to the STFT, the MDCT transformation amplitudes $A_k(m)$ are real-valued whereas the STFT amplitudes $\tilde{A}_k(m)$ are complex. The spectrogram of the STFT is defined as $|\tilde{A}_k(m)|^2$. Taking the spectrum of the STFT amplitudes removes the phase information of the original signal. Hence, generative models producing a spectrogram also need to generate the corresponding phase to produce the audio signal. The Griffin-Lim algorithm \cite{griffin} and other approaches exist to reconstruct the phase \cite{invphase, phaselearning}. Additionally, MDCT has strong energy compaction properties compared to other discrete transformations such as STFT \cite{DCTbook} meaning that a lower number of non-zero amplitudes are needed to carry the same amount of information. And finally, the MDCT representation allows to leverage the psychoacoustic filtering effect (see section \ref{pafilter}) 

MP3net uses the MDCT representation directly, hence not removing the implicit phase information and benefiting from the compaction of the audio signal in few amplitudes. See figure \ref{fig:spectrogram} for an example of the MDCT representation and the corresponding spectrogram. Note that we observed that it is beneficial for convergence and expressivity of the model not to convert the amplitudes to a (signed) log-scale (dB) when representing the audio signal in our model. Since MDCT is a linear transformation, superposition of two audio samples is achieved by adding the respective amplitudes. We hypothesize this helps the model to easily superimpose different sounds from different feature channels to generate the output. Another benefit is not switching to the (signed) dB is that it allows for the application of progressive training \cite{progan} to reduce the training time. Indeed, if one trains the deeper layers in the progressive training process, one needs to train the model on downsampled (blurred) data. However, if one were to blur the dB-rescaled amplitudes, the blurred data tends to average out at 0, leaving no information in the blurred sample (see section \ref{section:relatedwork}).

\subsection{Psychoacoustic filter} \label{pafilter}
\begin{figure}[ht]
    \centering
    \includegraphics[width=\columnwidth]{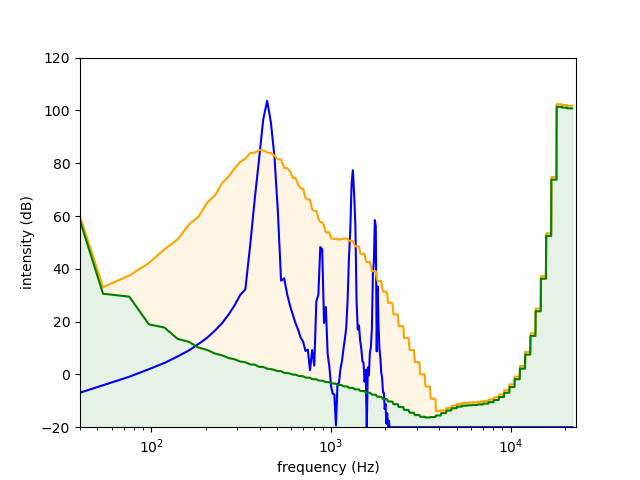}
    \caption{Example spectrum of a pure tone at 440Hz with 3 additional harmonics. The spectrum is plotted in blue. The green shading indicates the region of inaudible amplitudes due to the limit of perception of the human ear. Amplitudes in the area shaded in yellow are also inaudible since the are "over-powered" by the sound represented by the blue spectrum. Note that the first harmonic at 880Hz is entirely inaudible.}
    \label{fig:psychoacoutic_filter}
\end{figure}
The human ear is not able to distinguish between all the different possible raw audio signals. This phenomena is known as the psychoacoustic filter \cite{pafilterbook}. Mathematically this filter is expressed in the frequency spectrum, making the MDCT or STFT representation of the audio data the natural representation to apply this filter on.

The psychoacoustic filter consists of two key features. First, the filter leverages the fact that the human ear cannot hear certain quiet sounds. The level of perception depends on the frequency of the tone. Based on auditory experiments, one can approximate the amplitude intensity of this absolute threshold as \cite{pafilterbook}:
\begin{equation}
    I_{\text{absolute threshold}, j} = 10^{\frac{1}{10} L_{\text{absolute threshold}}(f_j)} 
\end{equation}
with 
\begin{multline}
    L_{\text{absolute threshold}}(f) [\mathrm{dB}] = \\ 3.64 f^{-0.8} - 6.5 e^{-0.6(f - 3.3)^2} + 10^{-3} f^4
\end{multline}
where $f$ is the frequency expressed in kHz. This function is commonly approximated as a stepwise function where $ L_{\text{absolute thresholdf}} $ is constant over a certain frequency range. Experimentally, these frequency ranges have been established as critical bands or Bark bands, with $f_j$ the mid point of each band. In figure \ref{fig:psychoacoutic_filter}, the absolute threshold is plotted in green. Amplitudes in the region below the absolute threshold (shaded in light green) are inaudible.

A second pyschoacoustic effect is that louder noises can render more quiet sounds inaudible, even if they are above the absolute threshold. This masking effect is also frequency dependent. The farther a frequency $j$ of an amplitude $A_j$ is from the frequency $i$ of the masking amplitude $A_i$, the weaker the masking effect:
\begin{equation} \label{eq:bark_start}
  I_{\text{mask}, j} = \left( \sum_i \left( A^{2}_i \right)^{\alpha} 10^{\frac{\alpha}{10} \left( f_{ij} - O_j \right)} \right)^{1 / \alpha}
\end{equation}
with $f_{ij}$ the spreading function and $O_j$ the offset:
\begin{eqnarray}
  f_{ij} & = & 15.81 + 7.5(i - j + 0.474) \nonumber \\ 
      & &  \quad - 17.5 \sqrt{1 + (i - j + 0.474)^2}\\
  O_j & = & \tau (14.5 + j) + (1 - \tau) 5.5
\end{eqnarray}
Here, $\alpha$ is a fixed non-linear superposition coefficient and $\tau$ is the tonality of the amplitude spectrum $A_k$ given by the spectral flatness measure:
\begin{equation} \label{eq:bark_end}
  \tau = \min \left(1, \frac{10}{-60\mathrm{dB}} \log_{10} \frac{ \exp \left( \frac{1}{N} \sum_k \ln A_k \right) }{ \frac{1}{N} \sum_k A_k} \right)
\end{equation}
which is $0$ for white noise and $1$ for a pure tone. The frequency indices in formulas (\ref{eq:bark_start})-(\ref{eq:bark_end}) refer to the mid-point frequencies of the respective Bark bands. In figure \ref{fig:psychoacoutic_filter}, the masking threshold of the spectrum (in blue) is plotted in yellow. Note that the first harmonic in the figure is below the masking threshold and hence inaudible. 

In MP3 compression, all amplitudes are quantized as integer multiples $a_k(m)$ of the auditory threshold:
\begin{equation}
  A_k(m) \approx a_k(m) \max(I_{\text{absolute threshold}, k}, I_{\text{mask}, k})
\end{equation}
This discretization process introduces an inaudible quantization error of size:
\begin{equation} \label{eq:quanterror}
  \frac{1}{2} \max(I_{\text{absolute threshold}, k}, I_{\text{mask}, k})
\end{equation}
In the MP3 format, the auditory threshold is stored for each Bark band, together with the integer multiples using Huffman coding. 

In generative models, this psychoacoustic auditive equivalence between two different audio signals can be leveraged to improve training stability and convergence. Alternatively it can be used to reduce the dimensionality of the problem. In MP3net, we opted to add gaussian noise with a standard deviation proportional to the psychoacoustic quantization error of equation (\ref{eq:quanterror}) to the MDCT amplitudes for both real and generated samples before they enter the discriminator. As studied by \citet{stablegan_withnoise} and \citet{stablegan_withnoise2}, this process improves the stability of the GAN training process, since it smooths the distribution of both the true and generated distributions, extending the support of both distributions. In earlier versions of our model, we implemented the psychoacoustic filter as a differentiable projection similar to a Leaky ReLU. This projection operator projected both real and generated samples onto a lower dimensional manifold by attenuating inaudible frequencies. However, the model yields better results with the gaussian noise approach.

We believe that the pyschoacoustic equivalence might also be useful for VAE-based generative models. The autoencoder can be trained to project out the psychoacoustic equivalence.

\section{Network architecture}
\subsection{2D convolutions to up- and downscale the MDCT amplitude representation}
Our network architecture is based on the architecture of the ProGAN network \cite{progan} with successive model blocks which scale up/down the image using strided 2D convolutions in the generator and discriminator respectively. Each model block in the generator takes the activation tensor from the previous layer as input and computes an activation tensor with double the resolution as output. The structure of the output tensor is such that each original input pixel is replaced with 2x2 pixels. Similarly, the model blocks of the discriminator halve the resolution of the input activation tensor. 

To build a similar convolutional network which doubles the audio resolution, we need to consider the two different ways we can increase the resolution given our MDCT amplitude tensor representation. First, we can double the sampling rate $f_s \rightarrow 2 f_s$ of $x(t)$ by doubling the number of blocks $M \rightarrow 2M$. Using the notation of equation (\ref{eq:sampling}), we get:
\begin{equation}
  \!\! x \! \left( \tilde{m}N t_0 + n \frac{t_0}{2} \right)
\end{equation}
with,
\begin{eqnarray}
  n & = & 0, \ldots, N-1 \\
  \tilde{m} & = & 0, \underbrace{\frac{1}{2}}_{\text{\ldots}}, 1, \ldots, \underbrace{M-\frac{3}{2}}_{N \text{ new blocks}}, M-1, \underbrace{M-\frac{1}{2}}_{\text{\ldots}}
\end{eqnarray}
We see that, similar to the image, each block $m$ of the input activation tensor $A_k(m)$ is replaced by two blocks $\tilde{m} = m$ and $m + 1/2$ in the output activation tensor $\tilde{A}_k(\tilde{m})$. We also note that while the number of filter bands $N$ has not changed, the corresponding frequencies of the filter bands have doubled to:
\begin{equation}
  \tilde{f}^0_k = \frac{(2f_s)}{2N} k \quad \text{with} \; k = 0, \dots, N - 1
\end{equation}

Alternatively, we can double the sampling rate of $x(t)$ by doubling the size of the blocks $N \rightarrow 2N$: 
\begin{equation}
  \!\! x \! \left( (m 2N + \tilde{n}) \frac{t_0}{2} \right) \text{with} \left\{
    \begin{array}{rcl}
      \tilde{n} & = & 0, \ldots, 2N-1 \\
      m & = & 0, \ldots, M-1
    \end{array}
  \right.
\end{equation}
This time there is no change to the block structure $m$ of the output activation tensor $\tilde{A}_{\tilde{k}}(m)$, but the filter bands corresponding with $\tilde{k}$ are now given by
\begin{equation}
   \tilde{f}^0_{\tilde{k}} = \frac{(2f_s)}{2(2N)} \tilde{k} 
\end{equation}
with,
\begin{equation}
  \tilde{k} = 0, \dots, N - 1, \underbrace{N, \dots, 2N - 1}_{N \text{ new output frequencies}}
\end{equation}
Here, we see that each frequency component $k$ is not replaced by two corresponding frequencies $\tilde{k} = k$ and $k + 1/2$, as was the case for the block structure $m$ when doubling the number of blocks. Instead, we first have the original $N$ frequencies from the input tensor and then we add $N$ new higher frequencies. Intuitively, we can understand this as adding a new higher-pitched octave with $N$ frequencies to our MDCT amplitude representation. Indeed, these new frequencies with even index $\tilde{k}$ are exactly double the frequencies of the highest octave already present in the input activation tensor:
\begin{eqnarray}
  \tilde{f}^0_{2k} &=& 2 f^0_k \; \text{where} \; k =  \underbrace{\frac{N}{2}, \dots, N - 1}_{N/2 \text{ input frequencies}}
\end{eqnarray}
Given this relation, we can take the MDCT amplitudes of the highest octave of the input tensor (frequencies $k = N/2, \dots, N-1$) and apply a transposed convolution with stride 2 to generate the new higher-pitched octave with $N$ amplitudes (frequencies $\tilde{k} = N, \dots, 2N - 1$). We then concatenate the newly generated octave with the input amplitudes.

\begin{figure}[t]
    \centering
    \includegraphics[width=\columnwidth]{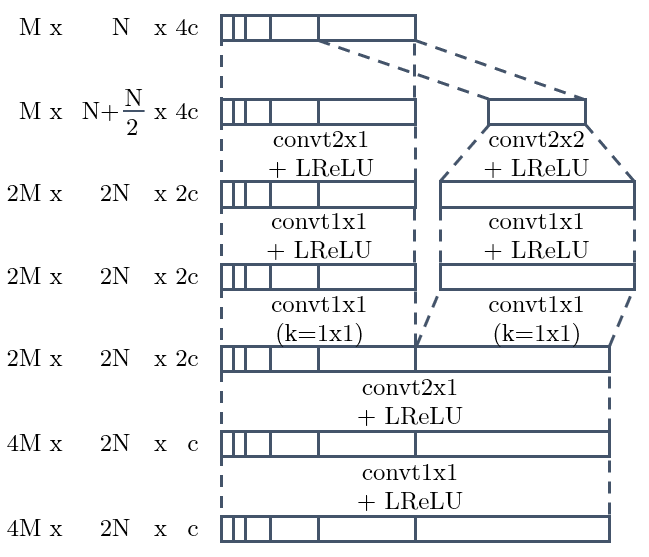}
    \caption{Design of the generator model block. The activation tensor of MDCT amplitudes $A_k(m)$ is represented by the horizontal bar, where the width illustrates the number of filter bands $k$. The (unbatched) dimensions of the tensors are indicated as $\text{block length} \times \text{filter bands} \times \text{channels}$. The notation for the transposed convolutions indicates the strides for the $\text{block} \times \text{filter band}$ directions.}
    \label{fig:modelblock}
\end{figure}
Using the above method we build a generator model block (see figure \ref{fig:modelblock}) which doubles the number of frequency components ($N \rightarrow 2N$) and quadruples the number of blocks ($M \rightarrow 4M$). To allow the model to balance the MDCT amplitudes of the newly created octave with the lower octaves, we introduce a convolutional layer with strides $1 \times 1$ and kernel $1 \times 1$ and without activation before concatenating.

The discriminator model block has the same structure as the generator model block, with the layer order reversed except for the position of the octave balancing layer. In the generator, this layer is located just before the downsampled highest octave is summed with the lower octaves.

\subsection{Overall model architecture}
\begin{figure}[ht]
    \centering
    \includegraphics[width=\columnwidth]{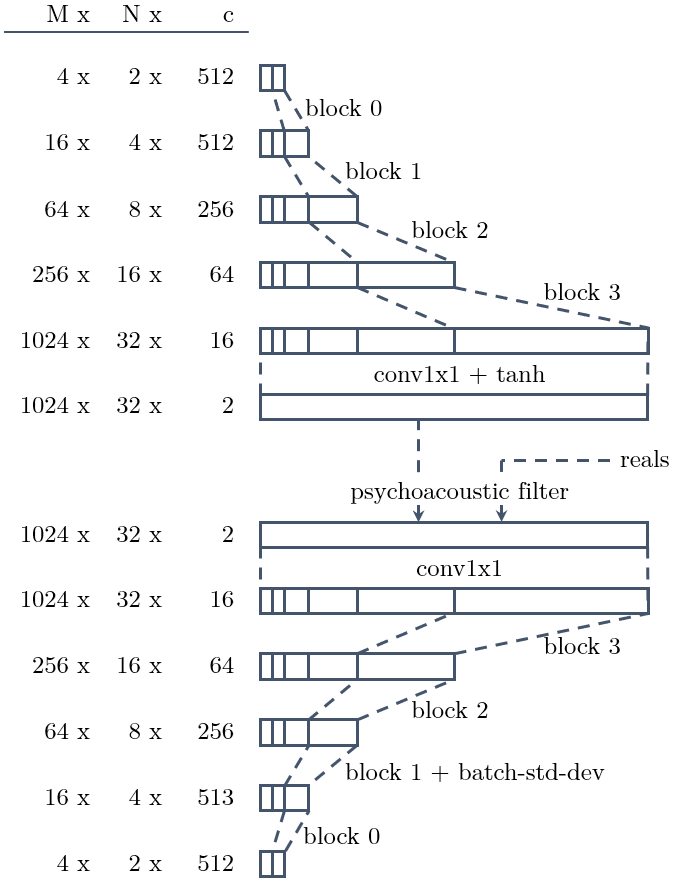}
    \caption{Model architecture of MP3net with 4 model blocks. The model to create 1m35s long samples with a 22,016Hz sample rate has 6 model blocks.}
    \label{fig:modelarchitecture}
\end{figure}
The model architecture of MP3net is shown in figure \ref{fig:modelarchitecture}. The model is a WGAN with gradient penalty. Similar to \citet{progan}, the discriminator loss function has a small drift term $\epsilon_{\text{drift}} \mathbb{E}_{x \in \mathbb{P}_r} [D^2(x)]$. 

Both real and generated MDCT amplitudes pass through the psychoacoustic layer where gaussian noise is added. The standard deviation of the gaussian is proportional to the quantization error computed in equation (\ref{eq:quanterror}).

The model also contains the minibatch standard deviation layer as described in \citet{progan}. This layer is added just before the last model block in the discriminator. It computes the variance over the minibatch for each of the elements in the activation tensor. This variance is added as an extra feature to the discriminator to force the generator to produce a diversity of generated samples. If the generator were only to produce a low diversity of samples, then the resulting low variance over the minibatch would help the discriminator to identify the samples as generated.

No batch normalization or pixelwise feature normalization layers (see \citet{progan}) are included in the generator. Note that we observed that standard batch normalization, where each feature is rescaled independently, destroys the training process as the model is unable to learn patterns which require calibration along the feature dimension.

The benefit of using a deep convolutional network is that since the deepest layers of our MP3net have the full context of the generated sample, they are able to produce samples with long-range coherence. Additionally, once the model is trained, inference is almost instantaneous. 

Given that 2D convolutional networks are well studied due to their importance in image generation, we can borrow some of the tools developed in that context to further improve model preformance. In particular progressive training of the deeper layers could significantly reduce the training time \cite{progan} and self-attention layers such as discussed in \cite{selfattention} can further improve long-range coherence of the generated audio sample.

\section{Experiments with the MAESTRO dataset}

\subsection{Dataset description}
Our experiments are based on the MAESTRO-V2.0.0 dataset \cite{maestro}. This dataset contains over 200h of classical piano music, recorded over nine years of the International Piano-e-Competition. To train MP3net, we use the WAV audio files from the competition years 2004, '06, '08 and '09 (51GiB of WAV files). These audio files were recorded with conventional recording equipment and hence contain some background noises. We resampled the audio to 22,016Hz and sliced the pieces into fixed-length audio samples for training. As such, the start of the training samples does not usually coincide with the start of the music piece. We did not condition MP3net on starting position of the sample in the song as was done in \citet{jukebox}

\subsection{Training details}
We ran two experiments. In the first experiment, the model generates 95-second audio samples to evalutate long-range coherence, musicality and diversity of the generated samples. In the second experiment, we reconfigure the model to produce 5-second samples to study audio quality and timbre.

\subsubsection{95-second model}
We used a model with a 512 dimensional latent space and 6 subsequent model blocks (see figure \ref{fig:modelblock}). The dimension of the generated output is 16,384 $\times$ 128 $\times$ 2 in the MDCT representation of $\text{block length} \times \text{filter bands} \times \text{channels}$. We capped the number of feature channels at 512 in the deeper layers of the network. The dimensions of our model was memory-bound by the 8GiB HBM of each TPUv2 core. We used a batch-size of 8 and split convolutions over activations with large block length into multiple convolutions along the batch dimension to avoid TPU padding overhead on the batch dimension. The generator and discriminator each have 59 million parameters. We used the Adam optmizer with learning rate $0.0001$,  $\beta_1 = 0.5$ and $\beta_2 = 0.9$. 

We trained the model for 250h on a single Cloud TPUv2, corresponding with 750,000 iterations of the discriminator (batch size 8), with two discriminator updates for each generator update. While we have not benchmarked the resulting generated samples of MP3net against other multi-minute generative audio models, such as jukebox \cite{jukebox}, the training time of MP3net is orders of magnitude smaller than jukebox which requires approximately 400,000h on a single V100\footnote{Depending on the model type and implementation details 1 to 3 NVIDIA V100s are roughly equivalent to 1 Cloud TPUv2, which has 8 cores \cite{hardwarebenchmarking}}. 

\subsubsection{5-second model}
\begin{figure*}[ht]
    \centering
    \begin{subfigure}[b]{\textwidth}
        \centering
    	\includegraphics[width=0.48\linewidth]{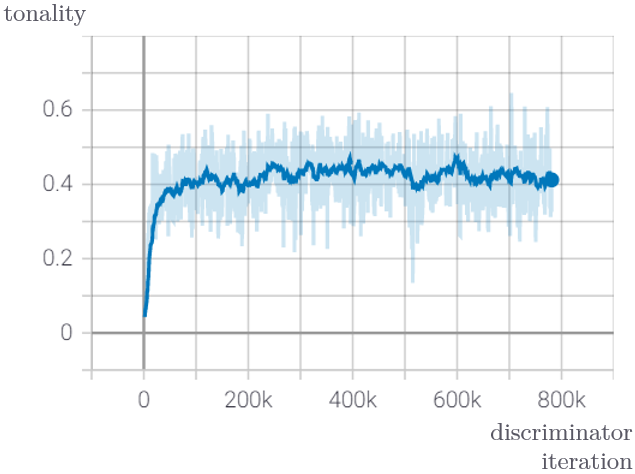}
        \hfill
        \includegraphics[width=0.48\linewidth]{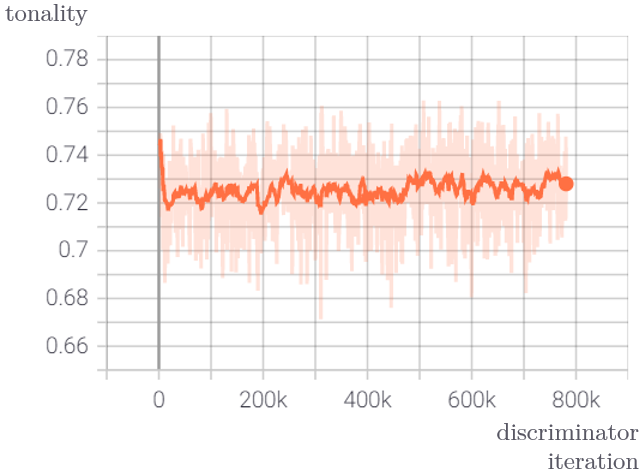}
    \end{subfigure}
    \caption{Evolution of the tonality as function of the discriminator iteration during the training process. The left chart shows the tonality of the generated samples. The right graph displays the tonality of the real samples used for training the model}
    \label{fig:tonality}
\end{figure*}
To study audio quality and timbre, we increased the number of features in the most shallow layer of the model to 128. We offset this increase in memory consumption by shortening the sample to 5s (1,024 $\times$ 128 $\times$ 2) with a model consisting of 5 subsequent model blocks. This model configuration has 64 million parameter for both the generator and the discriminator. We used the same Adam optimizer hyperparameters as for the 95s model and lowered $\beta_1$ to zero for the last 18,000 iterations to improve convergence.

The 5s model was trained for 120h on a single Cloud TPUv2, corresponding with 160,000 iterations of the discriminator (batch size 64).

\subsection{Observations}
A set of generated samples together with the source code of our model is available on the web page\footnote{\href{https://korneelvdbroek.github.io/mp3net/}{\url{https://korneelvdbroek.github.io/mp3net}}}.

\subsubsection{Audio quality and timbre}
A key parameter to compare the short-range performance of an audio model is the tonality $\tau = \mathbb{E}_m [\tau(m)]$ with $ \tau(m)$ defined in equation (\ref{eq:bark_end}). In figure \ref{fig:tonality}, we see that the tonality of the 95s samples quickly increases during initial training iterations, but then levels off in the 40-45\% range. Comparing with the real samples used for training which have a much higher tonality of around 72-73\% we see that a tonality gap of 30\% remains. The 5s samples exhibit a similar tonality gap.

Both the 95s model as well as the 5s model reproduce the piano timbre in the generated samples. However, the additional feature depth of the shallow layers of the 5s model results in a piano timbre which is clearly superior to that of the 95s model. The generated samples of the 5s model have a clearer and brighter piano sound, often with a better defined start of each note reproducing characteristic sound of the piano hammer hitting the string. Indeed, in the spectrogram of the 5s model one can more clearly identify the characteristic triangular structure of a piano note, where all harmonics start at exactly the same time and the higher harmonics fade out faster than the lower harmonics. The timbre of the 95s samples sometimes resembles more woodwind- or string-like timbres. 

During the training process, we often observe a characteristic humming sound both for the 5s and the 95s generated samples. The spectrograms of the samples with hum exhibit a checkerboard pattern with a periodicity along the block (time) direction linked with the more shallow model blocks. Similar checkerboard patterns have also been observed in images generated using (transposed) convolutions. In \citet{checkerboard}, this phenomena is described, together with techniques to avoid these artefacts. When freezing the weights of all but the two most shallow model blocks in both the generator and the discriminator, the tonality improves and the humming noise disappears after training for an additional 10-20k iterations. However, this comes at the expense of less interesting musicality in the generated samples. 

We also note that the expressivity of the highest octave in the samples produced by our model is limited. The model generates much less well defined harmonics in the highest octave, while many training samples do exhibit such structure in the highest octave.

\subsubsection{Musicality}
The rhythmic structure of the generated sample exhibits coherence from start to end of most 95s samples, with the tempo remaining constant throughout the sample.

The harmonic progressions in many of the generated samples follow the patterns common for western classical music. The chords structure is consistent between the start and end of these sample. Some other samples exhibit an atonal structure, without a clear tonal center. This latter is to be expected since ${\sim}4\%$ of the training dataset contains atonal pieces (mostly from Alexander Scriabin).
Some of the generated samples, exhibit clear melody. Even some of the 5s samples contain short musical phrases consisting of interesting motifs.

While the 95s clearly do not have the form and structure of a humanly composed piece, one can identify musical phrases starting softly (piano), building up with a crescendo towards a louder (forte) section. We have not identied any recurrent melodies (chorus) in any of the generated pieces.

\subsubsection{Sample diversity}
The MAESTRO dataset consists of music from the Baroque era over the Classical, Romantic and Impressionist styles, to the Expressionist period. Even though the generated samples are unprimed, samples resembling each of these styles are generated. The tempo of the generated samples also varies from slow pieces to samples with fast bravura. Harmony, chord progressions and melody are also very varied amongst the generated samples.

\section{Related Work} \label{section:relatedwork}
The field of synthetizing sound and music using computers is as old as computer science itself. With the advent of artificial neural networks new methods have become available to create audio. One of these techniques, Generative Adversarial Networks (GAN), uses two competing neural networks to generate a distribution of generated samples which closely approximates the true distribution of real samples. This technique was first described in the seminal paper by \citet{goodfellow2014}. 

A lot of research in recent years has contributed to improving the stability of the training process of GANs. \citet{wasserstein} introduced the WGAN model with a loss function based on the Wasserstein distance between distributions. This loss function helps to improve stability, in particular for distributions which have a low dimensional support compared to the full dimensionality of the data representation space. \citet{gradientpenalty} introduced a further key improvement to the WGAN technique by replacing the weight clipping with a gradient penalty in the loss function. Multiple other GAN flavors exist with different loss functions and other features to stabilize the training process \cite{zerogp, hingeloss, ganscan}. Recent research on GAN stability has identified the key role played by the singular values of the network kernels \cite{singularvalues}. 

GAN-based models have produced impressive results in the field of image generation. \citet{progan} introduced the ProGAN model. It can generate $1024 \times 1024$ images of faces in full color which are hard to distinguish from real pictures. The StyleGAN model \cite{stylegan}, is a further modification to the ProGAN model allowing one to tune the style of the generated image at each level of detail going from fine-grained features over middle-level styles, such as eyes, hair and lighting of the picture to high-level styles like hair style and face shape. The SAGAN model \cite{selfattention} uses a self-attention layer to increase the long-range coherence of convolutional neural nets and boost model performance for images which contain geometric structures. 
As these models become very large with millions to billions of parameters, new challenges in stabilizing the trainings process present themselves. \citet{biggan} explore these issues and give an extensive hyper-parameter scan for hinge-loss based model such as SAGAN.

Part of the research into audio generation has focus on symbolic music generation such as in \citet{deepbach} and \citet{musegan}. Direct generation of the raw audio is often more computationally intensive but required for key applications such as authentically reproducing human speech in text-to-speech. Several different neural network techniques have been applied to generate raw audio. WaveNet \cite{wavenet} uses an autoregression model to generate speech which mimics human voices. Using Variational Autoencoder (VAE) based model, Jukebox \cite{jukebox} produces impressive quasi-realistic multi-minute songs primed on music genre, artist and lyrics. GAN-based models on raw audio were introduced in WaveGAN \cite{wavegan}. 

Most closely related to the work presented here is GANsynth \cite{googlespecgan} and MelNet \cite{melnet}. GANsynth produces the STFT spectrograms for 4s audio samples. The GANsynth architecture is based on ProGAN. The resulting GANsynth audio samples of musical notes from different instruments are consistently judged by human evaluators of better fidelity compared to the similar samples generated by WaveNet. MetNet combines an RNN-based autoregressive model with a multi-scale generation procedure to generate STFT amplitude spectrograms which capture both the local as well as more long-range structures of spoken language and music.

\section{Future work}
\begin{figure}[ht]
    \centering
    \includegraphics[width=\columnwidth]{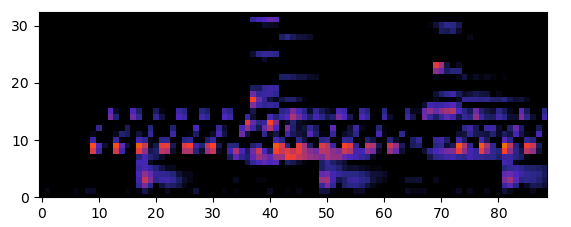}
    \caption{Reduced spectrogram of the first 10 bars Arvo P\"art's \textit{Spiegel im Spiegel}. The spectrogram corresponds closely to the score of the piece. The technique of reducing the spectrogram can potentially be used to speed up the training process with a progressive training approach.}
    \label{fig:blurring}
\end{figure}
Since the model architecture of MP3net is very similar to the convolutional networks well studies in the field of image generation, we can borrow many of the techniques of image generation. In particular, using progressive training \cite{progan} we might be able to reduce training time significantly and improve convergence. Progressive training, leverages the layered structure of the model where deeper layers are trained first. This training process is much faster since the memory footprint of the features in the deeper layers is much smaller allowing for larger batch sizes and since less weights have to be optimized. 

To apply progressive training, one needs to guess (or impose) how the data is represented in the deeper layers. In the case of images, the authors of ProGAN  take blurred versions of the real images and present these when training the deep layers. While the MP3net architecture is similar in many respects to ProGAN, the characteristics of the data representation is rather different. Images consist of regions where adjacent pixels vary smoothly, where our MDCT amplitude representation of audio signals exhibits periodic functions that oscillate between -1 and 1 (see figure \ref{fig:spectrogram}). 

Initial experiments show that if we were to rescale the MDCT amplitudes to the dB-scale as is customary, blurring the images would very quickly remove all the information content from the image as the oscillating amplitudes would cancel out to zero. Our non-rescaled amplitudes don't exhibit this disadvantage. Experimentation with blurring the highest octave and subsequently folding (summing) it with the second highest octave has the advantage that harmonics in different octaves are summed together. Repeating this procedure gives a reduced spectrogram which could be used as the training data for the deeper layers. Visually, such a reduced spectrogram resembles music scores since harmonics are reduced/folded on top of the fundamental frequency. Note that this technique to reduce the spectrogram does not give the same results as merely downsampling the audio and then converting to a spectrogram. Figure \ref{fig:blurring} shows an example of a reduced spectrogram where the pixels that light up correspond clearly to the notes of the music piece. 

Another prerequisite to successfully apply progressive training is that the data representation in each of the layers of the generator is properly normalized. Often, this is accomplished with batch normalization. In \citet{progan}, the authors apply a pixelwise feature normalization to obtain the same result. 

We would like to explore the long-range coherence on a more divers set of training data. In particular, training on the Blizzard \cite{blizzard2011} and VoxCeleb2 datasets \cite{voxceleb2} would test the model's ability to synthesis speech. Transformer techniques as used in \citet{selfattention} can further improve the long-range coherence of the generated samples. 

Another direction to explore is conditioning the network. \citet{melnet} conditioned MelNet on text for text-to-speech synthesis. In \citet{jukebox}, the model is conditioned on the lyrics of songs in a lyrics-to-song synthesis. Another approach to controlling the generated samples is to upgrade the network structure in line with StyleGAN \cite{stylegan}. This modified architecture would allow us to control the generated audio at different scales.

\section{Conclusion}
In this paper, we introduce MP3net, a 2D convolutional GAN with an architecture similar to some of the most successful image generation GANs \cite{progan, selfattention, stylegan}. MP3net borrows techniques from the field of audio compression. In particular, we use the MDCT as data representation since it includes all phase information of the original audio sample and we leverage the pyschoacoustic properties of human ears to simplify the training problem by widening our data and generated distributions. 
Our model is able to produce high-quality, stereo audio with relative limited computational power. The samples produced exhibit long-range coherence over the full 95s of the samples produced. This ability is explained in part by the fact that the deepest convolutional layers of the model do have the full context of the sample. Hence they are able to generate music features that are coherent from the start to the end of the sample. 
Compared to other generational models generating multi-minute samples, training times of MP3net are much shorter and inference is quasi-instantaneous given the inherent CNN-model architecture.

\section{Acknowledgement}
We would like to thank Werner Van Geit for providing feedback on the initial draft of this paper.
We also extend gratitude to Google Colab for making powerful compute available for everyone.

\bibliography{main}

\bibliographystyle{icml2020}

\end{document}